\documentclass[12pt]{article}
\usepackage{epsfig}
\parskip=\itemsep               
\date{}

\catcode`@=11

\def\@citex[#1]#2{\if@filesw\immediate\write\@auxout{\string\citation{#2}}\fi
  \def\@citea{}\@cite{\@for\@citeb:=#2\do
    {\@citea\def\@citea{,\penalty\@m}\@ifundefined
      {b@\@citeb}{{\bf ?}\@warning
       {Citation `\@citeb' on page \thepage \space undefined}}%
\hbox{\csname b@\@citeb\endcsname}}}{#1}}

\def\citer{\@ifnextchar [{\@tempswatrue\@citexr}{\@tempswafalse\@citexr[]}}

\def\@citexr[#1]#2{\if@filesw\immediate\write\@auxout{\string\citation{#2}}\fi
  \def\@citea{}\@cite{\@for\@citeb:=#2\do
    {\@citea\def\@citea{--\penalty\@m}\@ifundefined
       {b@\@citeb}{{\bf ?}\@warning
       {Citation `\@citeb' on page \thepage \space undefined}}%
\hbox{\csname b@\@citeb\endcsname}}}{#1}}
\catcode`@=12

\def\bo{{\raise.15ex\hbox{\large$\Box$}}}               
\def\face{{\raise.2ex\hbox{$\displaystyle \bigodot$}\mskip-2.2mu \llap {$\ddot
        \smile$}}}                                      

\def\leftrightarrowfill{$\mathsurround=0pt \mathord\leftarrow \mkern-6mu
        \cleaders\hbox{$\mkern-2mu \mathord- \mkern-2mu$}\hfill
        \mkern-6mu \mathord\rightarrow$}       
\def\dvec#1{\vbox{\ialign{##\crcr
        \leftrightarrowfill\crcr\noalign{\kern-1pt\nointerlineskip}
        $\hfil\displaystyle{#1}\hfil$\crcr}}}           

\def\beq{\begin{equation}}
\def\eeq{\end{equation}}

\def\beqx{\begin{displaymath}}
\def\eeqx{\end{displaymath}}

\def\beql{\begin{eqnarray}}
\def\eeql{\end{eqnarray}}

\def\Journal#1#2#3#4{{#1} {\bf #2} (#4) #3}

\def\NPB{{\em Nucl. Phys.} B}
\def\PLB{{\em Phys. Lett.} B} 
\def\PRL{\em Phys. Rev. Lett.} 
\def\PRD{{\em Phys. Rev.} D} 
\def\ZPC{{\em Z. Phys.} C} 
\def\PR{\em Phys. Rev.}

\def\PR{\em Phys. Rep.}

\newcommand{\lwig}{\mbox{\,\raisebox{.3ex}
    {$<$}$\!\!\!\!\!$\raisebox{-.9ex}{$\sim$}\,}}
\newcommand{\gwig}{\mbox{\,\raisebox{.3ex}
    {$>$}$\!\!\!\!\!$\raisebox{-.9ex}{$\sim$}}\,}

\newcommand{\ai}{{\overline{I}}} 
\newcommand{\iai}{I\overline{I}}

\newcommand{\ii}{{\rm i}} 
 
\newcommand{\xpr}{{x^\prime}}

\newcommand{\ts}{\tilde{S}}
\newcommand{\dts}{D(\tilde{S})}
\newcommand{\ddts}{D\left( \ln \left( \frac{\dts}{\sqrt{\xi^\ast -2}}
                   \right)\right)}

\begin{document}
\title{
{\normalsize\rightline{DESY 98-081}\rightline{hep-ph/9806528}} 
\vskip 1cm 
      \bf Instanton-Induced Cross-Sections\\ 
          in Deep-Inelastic Scattering
                          \\
  \vspace{11mm}}
\author{A. Ringwald and F. Schrempp\\[4mm] 
Deutsches Elektronen-Synchrotron DESY, Hamburg, Germany}
\begin{titlepage} 
  \maketitle
\begin{abstract}
We present our results for  inclusive instanton-induced   
cross-sections in deep-inelastic scattering, paying in particular attention to
the residual renormalization-scale dependencies. A ``fiducial'' kinematical 
region in the relevant Bjorken variables is extracted from recent
lattice simulations of QCD. The integrated instanton-contribution to the 
cross-section at HERA corresponding to this fiducial
region is surprisingly large: It is in the ${\mathcal O}(100)$ pb range, and 
thus remarkably close to the recently published experimental upper bounds.
\end{abstract} 
\thispagestyle{empty}
\end{titlepage}
\newpage \setcounter{page}{2}


{\bf 1.}
Instantons~\cite{bpst} are non-perturbative gauge field fluctuations. They 
describe {\it tunnelling} transitions between degenerate vacua 
of {\it different topology} in non-abelian gauge theories like QCD. 
Correspondingly, in\-stan\-tons and anti-instantons carry an 
{\it integer topologigal} charge $Q =1$ and $Q=-1$, respectively, while the 
usual perturbation theory resides in the sector $Q=0$. Unlike the latter, 
instantons induce processes which
violate {\it chirality} ($Q_5$) in (massless) QCD, in accord~\cite{th} with 
the general chiral-anomaly relation. An experimental discovery of 
instanton-induced events would clearly be of basic significance.

The deep-inelastic regime is distinguished by the fact that here
hard in\-stan\-ton-induced processes may both be {\it
calculated}~\cite{bb,mrs1} within instanton-per\-tur\-ba\-tion
theory  and possibly 
{\it detected experimentally}~\cite{rs,grs,rs2,crs}.
As a  key feature it has recently been shown~\cite{mrs1}, that in 
deep-inelastic scattering (DIS) the 
generic hard scale ${\cal Q}$ cuts off instantons with {\it large size} 
$\rho\gg {\cal Q}^{-1}$, over which one has no control theoretically. 

In continuation of Ref.~\cite{mrs1}, where the amplitudes and cross-sections 
of {\it exclusive} partonic subprocesses relevant for DIS were calculated, we 
summarize in the present letter the results of our finalized
calculations of the various {\it inclusive} instanton-induced  
cross-sections (Sects.~2 and 4). A 
detailed account of our calculations will be published elsewhere~\cite{mrs3}.
The essential new aspect as compared to our first estimates~\cite{rs2} is the 
strong reduction of the residual dependence on the renormalization scale 
resulting from a recalculation based on an improved instanton 
density~\cite{morretal}, which is renormalization-group invariant at the 
two-loop level.

There has been much recent activity in the lattice community 
to ``measure'' topological fluctuations in 
lattice simulations~\cite{lattice} of QCD. Being
independent of perturbation theory, such simulations provide
``snapshots'' of the QCD vacuum including all possible
non-perturbative features like instantons. They also provide crucial 
support for important prerequisites of our calculations in DIS,
like the validity of instanton-perturbation theory and 
the dilute instanton-gas approximation for {\it small} instantons of size
$\rho \leq {\mathcal O}(0.3)$ fm. As a second main point of this letter 
(Sect.~3), these lattice constraints will be exploited and translated into a 
``fiducial'' kinematical region for our predictions of the
instanton-induced DIS cross-section based on instanton-perturbation theory.

\begin{figure}[ht]
\vspace{-0.2cm}
\begin{center}
\epsfig{file=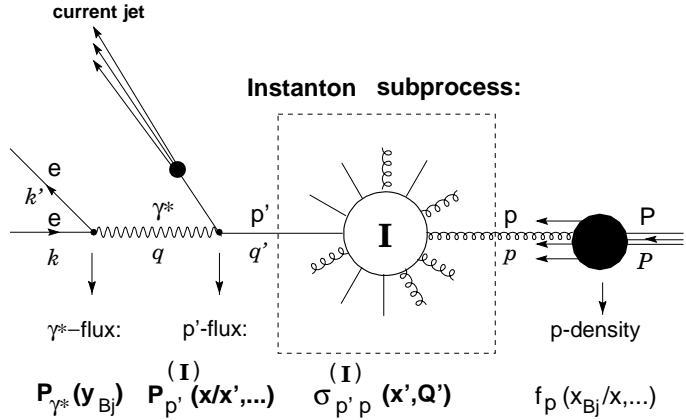,width=9cm}
\caption[dum]{\label{ev-displ}The leading instanton-induced process 
in the DIS regime of $e^\pm P$ scattering.}
\end{center}
\end{figure}

{\bf 2.}
The leading instanton ($I$)-induced process in the DIS regime of $e^\pm
P$ scattering for large photon virtuality $Q^2$ is illustrated in 
Fig.~\ref{ev-displ}. The dashed box emphasizes the  so-called  
instanton-{\it subprocess} with its own Bjorken variables,
\begin{equation}
        Q^{\prime\,2}=-q^{\prime\,2}\geq 0;\hspace{0.5cm}
        \xpr=\frac{Q^{\prime\,2}}{2 p\cdot q^\prime}\le 1.
\end{equation}
As can be inferred from Ref.~\cite{rs2} and will be detailed in 
Ref.~\cite{mrs3}, the inclusive $I$-induced  
cross-section\footnote{A sum over $I$-induced
($\triangle Q_5=2n_f$) and anti-instanton ($\ai$)-induced 
($\triangle Q_5=-2n_f$) processes is always implied by the superscript $(I)$ 
at cross-sections.} in unpolarized deep-inelastic $e^\pm P$ scattering can be 
expressed (in the Bjorken limit) as
\begin{equation}
\label{ePcross}
\frac{d\sigma_{eP}^{(I)}}{d\xpr\,dQ^{\prime 2}} \simeq
\sum_{p^\prime,p}
\frac{d{\mathcal L}^{(I)}_{p^\prime p}}{d\xpr\,dQ^{\prime 2}}
\,\sigma_{p^\prime p}^{(I)}(\xpr,Q^{\prime 2}) ,
\end{equation} 
where $p^\prime =q^\prime,\overline{q}^\prime$ denotes the virtual 
(anti-)quarks entering the $I$-subprocess from the photon side and
$p=q,\overline{q},g$ denotes the target partons (c.f. Fig.~\ref{ev-displ}). 
The differential luminosity $d{\mathcal L}^{(I)}_{p^\prime p}$, accounting
for the number of $p^\prime p$ collisions per $eP$ collision, has a 
convolution-like structure~\cite{rs}, involving 
$\{x_{\rm Bj},y_{\rm Bj},x\}$-integrations over the target-parton density,
$f_p(x_{\rm Bj}/x,\dots)$, the $\gamma^\ast$-flux,
$P_{\gamma^\ast}(y_{\rm Bj})$, and the {\it known}~\cite{rs2,mrs3} 
flux $P^{(I)}_{p^\prime}(x/\xpr,\ldots )$ of the parton $p^\prime$ in the 
$I$-background. We shall display the explicit form of the differential 
luminosity in Sect.~4. 

The simple relation (\ref{ePcross}) between $d\sigma_{eP}$
and $\sigma_{p^\prime p}^{(I)}$, derived
within $I$-perturbation theory~\cite{rs2,mrs3} in the Bjorken limit, 
is actually much less obvious than an inspection of the grossly oversimplified
Fig.~\ref{ev-displ} may suggest. The derivation proceeds in two steps:  
$i)$ Using the Feynman rules of $I$-perturbation theory in momentum space,
one calculates the (manifestly gauge invariant)
inclusive $e P$ cross-section, with the current quark being not a free
parton, but rather described by the (complicated) quark propagator in the
$I$-background as in Ref.~\cite{mrs1}. $ii)$ One independently
writes down  the (gauge invariant) expression for the total 
cross-section $\sigma_{p^\prime  
p}^{(I)}$ with an {\it off-shell} external parton $p^\prime$.       
In the Bjorken limit, when certain non-planar contributions may be
neglected, one then finds by comparison of $i)$ and $ii)$ the 
form (\ref{ePcross}) of the inclusive $I$-induced
 $eP$ cross-section along with the explicit expression for
the flux-factor $P_{p^\prime}^{(I)}$.

While a more detailed description of this rather involved calculation
has to be deferred elsewhere~\cite{mrs3}, let us summarize next the
state of the art evaluation of the $I$-subprocess total 
cross-section $\sigma_{p^\prime
p}^{(I)}(\xpr,Q^{\prime 2})$, which contains most of the crucial 
instanton-dynamics.  

We start with the $I$-subprocess total cross-sections (here only 
for the dominating case of a target gluon)
in a form~\cite{rs2,mrs3} still exhibiting the 
complicated integrations over collective coordinates\footnote{For brevity, we 
display the cross-section (\ref{qgforwampl}) already after the (saddle-point)
integration over the $I$-colour orientations. The function 
$\tilde\Omega$, whose explicit form will be specified below, accounts for
that.} ($I$-sizes 
$\rho,\overline{\rho},\ldots$), 
\begin{eqnarray}
\sigma_{p^\prime g}^{(I )}
 &\simeq& 
\int\limits_{0}^{\infty} d\rho
\int\limits_{0}^{\infty} d\overline{\rho}
\int d^4 R\,\,D(\rho; \mu_r ) \,D(\overline{\rho};\mu_r) \, 
\left(\rho\overline{\rho}\mu_r^2\right)^{
\beta_0\Delta_1\,
\Omega\left(\frac{R^2}{\rho\overline{\rho}},
\frac{\overline{\rho}}{\rho}\right)}
\,
\nonumber
\\[1.6ex]
\label{qgforwampl}
&&\times \,
K_1\left( Q^\prime\,\rho\right)
       K_1\left( Q^\prime\,\overline{\rho}\right)\,
\exp\left[ \ii\,(p+q^\prime)\cdot R\right]\,
\exp\left[ -\frac{4\pi}{\alpha_s(\mu _r )}\,
\Omega\left(\frac{R^2}{\rho\overline{\rho}},
\frac{\overline{\rho}}{\rho}\right) 
     \right]
\\[1.6ex]
\nonumber
&&\times \,
\frac{1}{9\sqrt{\pi}}\, 
\left( \frac{\alpha_s(\mu_r)}{4\pi}
\frac{6}{{\tilde\Omega\left(\frac{R^2}{\rho\overline{\rho}},
\frac{\overline{\rho}}{\rho}\right)}}  \right)^{7/2}
\left( \rho\overline{\rho} \right)^{9/2} \,
\left[\omega\left(\frac{R^2}{\rho\overline{\rho}},
\frac{\overline{\rho}}{\rho}\right)\right]^{2\,n_f-1}
\frac{2}{3}\,\frac{\pi^5}{\alpha_s(\mu_r)}\,
\frac{Q^{\prime 4} (p\cdot q^\prime )}{((p+q^\prime )^2)^{3/2}}\,
\, ,
\end{eqnarray}
where $p^\prime =q^\prime,\overline{q}^\prime$.
The most important quantities entering Eq.~(\ref{qgforwampl}) are:
\begin{itemize}
\item {\it The $I$-density $D(\rho ,\mu_r)$,} 
      which has the general form~\cite{th,ber}
      \begin{equation}
      D(\rho, \mu_r) =
      d \left(\frac{2\pi}{\alpha_s(\mu_r)}\right)^6
      \exp{\left(-\frac{2\pi}{\alpha_s(\mu_r)}\right)}
      \frac{(\rho\, \mu_r)^{\beta_0\Delta_1-\Delta_2}}{\rho^{\,5}},
      \label{density} 
      \end{equation}
      with $\mu_r$ denoting the renormalization scale and~\cite{morretal} 
      \begin{equation}
      \label{Deltas}
      \Delta_1\equiv 1+\frac{\beta_1}{\beta_0}\frac{\alpha_s(\mu_r)}{4\pi};
      \hspace{6ex} 
      \Delta_2\equiv 12\,\beta_0\frac{\alpha_s(\mu_r)}{4\pi} ,
      \end{equation}
      in terms of the QCD $\beta$-function coefficients,
      $\beta_0=11-\frac{2}{3}{n_f};\ \beta_1=102-\frac{38}{3} {n_f}$.
      The power $\beta_0\Delta_1-\Delta_2$ makes the 
      $I$-density renormalization-group invariant at the two-loop 
      level~\cite{morretal}, 
      $(1/D){\rm d}D/{\rm d}\mu_r={\mathcal O}(\alpha_s^2)$, in contrast to
      the original one-loop expression~\cite{th}, corresponding to 
      $\Delta_1=1$ and $\Delta_2=0$, with 
      $(1/D){\rm d}D/{\rm d}\mu_r={\mathcal O}(\alpha_s)$.  
      The constant $d$ is scheme-dependent; in the 
      $\overline{\rm MS}$-scheme
      it is given by~\cite{dMS}
      $d=C_1 \exp [-3\,C_2+n_f\,C_3]/2$,
      with $C_{1}=0.46628$, $C_{2}=1.51137$, and $C_{3}=0.29175$.

      Note that the large, positive power of $\rho$ in the 
      $I$-density~(\ref{density}) would make the integrations over 
      the $I$-sizes in Eq.~(\ref{qgforwampl}) infrared divergent without
\item {\it the form factors $K_1(Q^\prime \rho (\overline{\rho}))$:}
      For large $Q^{\prime}\rho (\overline{\rho})$,       
      the virtuality $Q^{\prime}$ of the internal quark $p^\prime$ in 
      Fig.~\ref{ev-displ} provides an exponential cut-off, 
      $K_1 (Q^\prime\rho ) \propto\exp(-Q^\prime\rho)$, in the 
      integrations over the $I$-sizes~\cite{mrs1}.
      These form factors where shown to arise naturally in step  
      {\it i)} above, which is manifestly gauge invariant~\cite{mrs1}.
      In step {\it ii)} one has to adopt a gauge-invariant definition
      of the $p^\prime p$ cross-section, since the incoming parton
      $p^\prime$ is {\it off-shell}~\cite{b}. Then one obtains exactly the 
      Bessel-K form factors~\cite{mrs3}, unlike naive, not manifestly 
      gauge-invariant definitions which lead in addition to these well-defined 
      contributions to unphysical ones suffering from infrared divergent 
      $I$-size integrations~\cite{valley-most-attr-orient,bbgg}.
\item {\it The functions $\Omega$ and $\omega$} (along with the integration
      over $R_\mu$) 
      summarize the effects of final-state gluons ($\Omega$) and final-state 
      quarks ($\omega$). The function $\Omega$, appearing in the 
      exponent with a large numerical coefficient, $4\pi/\alpha_s$, and 
      $\omega$, occuring with a high power, $2n_f-1$, call for a precise 
      evaluation. Hence, let us turn next to describing their state-of-the-art
      evaluation. 
\end{itemize}

It is very instructive to consider two alternative interpretations of the 
functions $\Omega$, $\omega$, and the integration variable $R_\mu$.
\begin{itemize} 
\item {\it Total cross-section via summation of exclusive cross-sections:} 
  
      This is the cleanest and most straightforward method to arrive
      at the total cross-section. 
      In this case one starts with the familiar representation of the 
      $\delta^{(4)}$-function associated with energy-momentum
      conservation,
      \begin{equation}
      (2\,\pi)^{4}\, \delta^{(4)}( p+q^\prime -\sum_i k_i)
      =\int d^4R\ \exp\, [ \ii\,
      ( p+q^\prime -\sum_i k_i)\cdot R ].
      \end{equation}
      The phase-space integration over the final-state gluons/quarks is
      then performed by means of the basic formula
      \begin{equation}
      \int\frac{d^{4}k_{i}}{(2\,\pi)^{3}}\, 
      \delta^{(+)}( k_{i}^{2})\,\exp\left[ -\ii\,k_i\cdot R\right]
      =\frac{1}{(2\pi)^2}\,\frac{1}{-R^2 +\ii\epsilon R_0} ,
      \end{equation}
      with the help of which one finds~\cite{holypert}
      \begin{eqnarray}
      \label{action-asy}
      \Omega\left(\frac{R^2}{\rho\overline{\rho}},
      \frac{\overline{\rho}}{\rho}\right) &=& 
      - 6\left(\frac{\rho\overline{\rho}}{-R^2+\ii\epsilon R_0}\right)^2
      +12\left(\frac{\rho\overline{\rho}}{-R^2+\ii\epsilon R_0}\right)^3
      \left(\frac{\rho}{\overline{\rho}}+\frac{\overline{\rho}}{\rho} \right)
      +\ldots
      \\  
      \label{omega-asy}
      \omega\left(\frac{R^2}{\rho\overline{\rho}},
      \frac{\overline{\rho}}{\rho}\right) &=&
      4\left(\frac{\rho\overline{\rho}}{-R^2+\ii\epsilon R_0}\right)^{3/2}
      +\ldots
      .
      \end{eqnarray}
      The interpretation of the various terms contributing to the perturbative 
      expansion of $\Omega$ in
      Eq.~(\ref{action-asy}) is illustrated in Fig.~\ref{fhgpert} (left): The 
      first term takes into account the summation and exponentiation of the 
      leading-order gluon emission, whereas the 
      second term originates from the summation and exponentiation of 
      interference terms between the leading-order gluon emission and the 
      gluon-propagator correction. The first term contributing to the 
      perturbative expansion of $\omega$ in Eq.~(\ref{omega-asy}) 
      just corresponds to the leading-order quark emission. 
\begin{figure}[h]
\vspace{-0.4cm}
\begin{center}
\parbox{8.2cm}{\epsfig{file=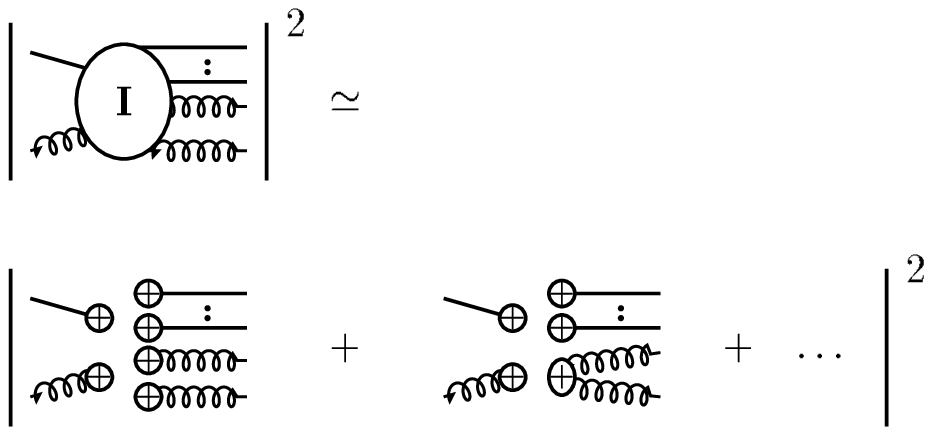,width=8cm}}\hspace{2.5ex}
\parbox{8.2cm}{\epsfig{file=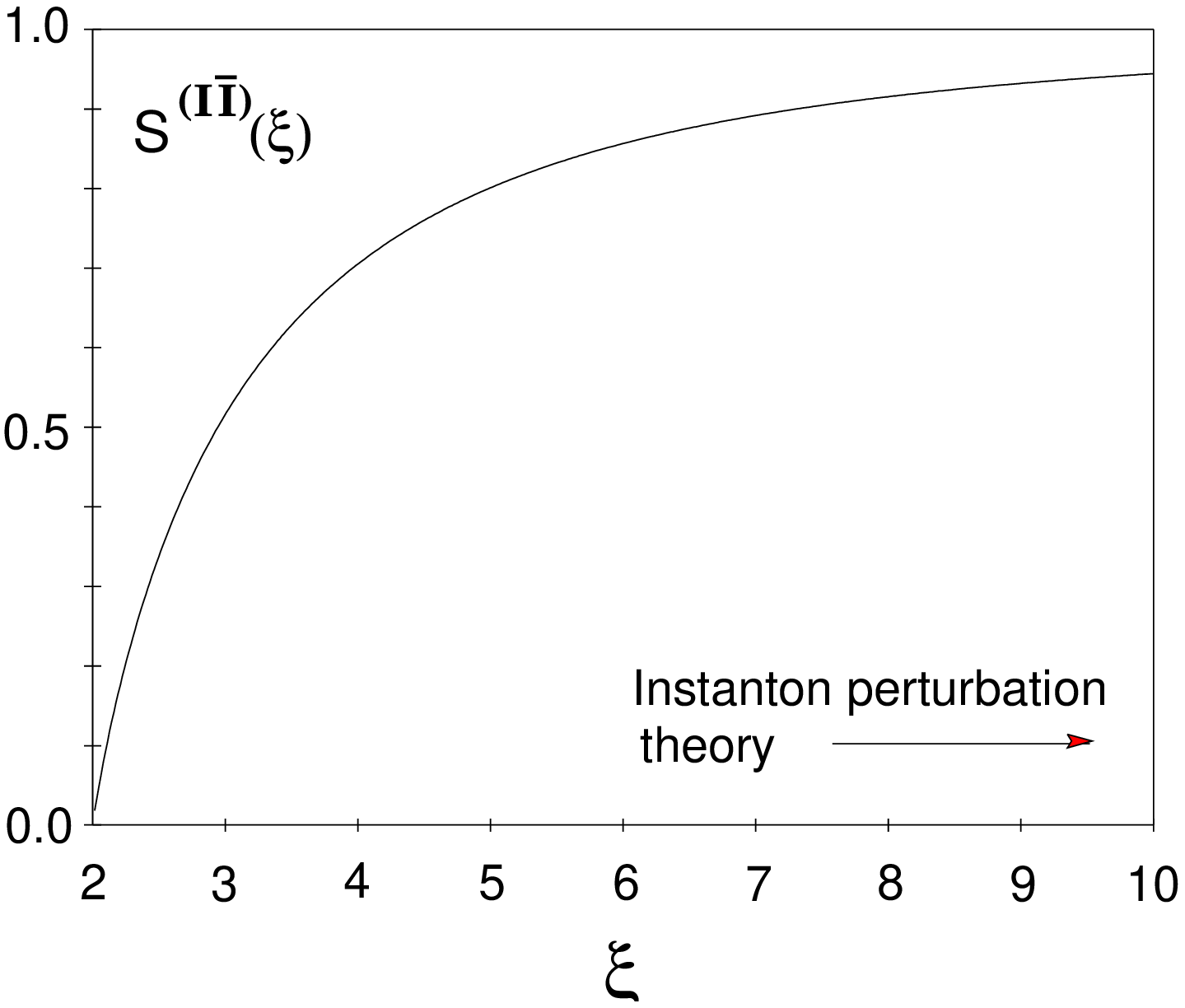,width=6.5cm}}
\caption[dum]{\label{fhgpert}
{\it Left:} Graphical interpretation of an exclusive $I$-induced $q^\prime g$ 
            cross-section in terms of perturbation theory in the 
            $I$-background. (Curly) lines ending at blobs denote LSZ-amputated
            quark zero modes (classical $I$-gauge
            fields). Curly lines connected by an ellipse denote LSZ-amputated
            gauge-field propagators in the $I$-background.
{\it Right:} The $\iai$-valley action corresponding to the most
attractive $\iai$-colour orientation.
}
\end{center}
\end{figure}
      
     
      In summary, the perturbative approach based on the exclusive 
      amplitudes, as calculated within $I$-perturbation theory, yields
      the essential functions $\Omega$ and $\omega$ as 
      asymptotic expansions for small $\rho\overline{\rho}/R^2$.
      Since $\rho$ and $\overline{\rho}$ are conjugate to the virtuality
      $Q^\prime$ and $R$ is conjugate to the total momentum of the 
      $I$-subprocess, $p+q^\prime$, (c.f. Eq.~(\ref{qgforwampl}) and 
      Fig.~\ref{ev-displ}), we expect qualitatively 
      \begin{equation}
      \label{qual-exp}
      \rho \sim \overline{\rho}\sim 1/Q^\prime \ {\rm and}\  
      R^2\sim 1/(p+q^\prime)^2 \ \Rightarrow \ 
      \rho\overline{\rho}/R^2 \sim 
      (p+q^\prime)^2/Q^{\prime 2}=1/\xpr -1.
      \end{equation} 
      Thus, strict $I$-perturbation theory for
      the total cross-section is only applicable for not too small $\xpr$. 
  
\item {\it Total cross-section via optical theorem and $\iai$-valley method:}

      In this approach, one evaluates~\cite{optvalley} the total 
      cross-section from the imaginary part of the forward elastic scattering 
      amplitude induced by the instanton-anti-instanton ($\iai$)-valley 
      background, $A_\mu^{(\iai)}$. In this case, $R_\mu$ stands for the 
      {\it separation} between $I$ and $\ai$, and $\Omega$ is identified with 
      the  
      {\it interaction} between $I$ and $\ai$,
      \begin{equation}
      \Omega \simeq S^{(\iai)}(\xi)-1.
      \label{interaction}
      \end{equation}
      In the valley approximation, the $\iai$-valley action,
      $S^{(\iai)}\equiv \frac{\alpha_s}{4\pi}\,S[A_\mu^{(\iai)}]$, 
      is restricted by conformal
      invariance to depend only on  the ``conformal separation''~\cite{yung}
      \begin{equation}
      \xi \equiv \frac{-R^2+\ii\epsilon R_0}{\rho\overline{\rho}}
      +\frac{\rho}{\overline{\rho}}+\frac{\overline{\rho}}{\rho} ,
      \end{equation}  
      and its functional form is explicitly 
      known~\cite{valley-most-attr-orient,valley-gen-orient}
      (Fig.~\ref{fhgpert} (right)).
      
      Note that for all separations $\xi$, the interaction between $I$
      and $\ai$ is {\it attractive} 
      (c.f. Fig.~\ref{fhgpert} (right)): The $\iai$-valley corresponds to
      a configuration of steepest descent interpolating between an infinitely 
      separated $I$/$\ai$ pair and a strongly overlapping one, annihilating to
      the perturbative vacuum.  

      Analogously, the function $\omega$ is now identified with
      the {\it fermionic overlap integral} for which an integral 
      representation was found in Ref.~\cite{shuverb}, which we were able to 
      perform analytically,
      \begin{equation}
      \label{exactom}
      \omega(\xi)=
      \frac{6 B(\frac{3}{2},\frac{5}{2})}{z^{3/2}}\,
      \phantom{}_2F_1\left(\frac{3}{2},\frac{3}{2};4;1-\frac{1}{z^2}\right);
      \hspace{6ex}
      z \equiv \frac{1}{2}\left(\xi +\sqrt{\xi^2-4}\right).
      \end{equation}

      Finally, the function $\tilde\Omega$ arising from the integration
      over the relative $\iai$-colour orientations has been estimated in
      Ref.~\cite{bbgg} by assuming for simplicity an orientation dependence of
      the valley 
      action\footnote{The saddle-point corresponds to the most-attractive 
      $\iai$-orientation. We have checked~\cite{mrs3} that taking into
      account the 
      {\it exact} orientation dependence of the valley 
      action~\cite{valley-gen-orient} 
      gives {\it numerically} a very similar result.} corresponding to a 
      dipole-dipole interaction~\cite{cdg}, 
      \begin{equation}
      \label{orient-valley}
      \tilde{\Omega}(\xi )\simeq \xi\,\frac{d\Omega (\xi)}{d\xi} .
      \end{equation}  

      The {\it leading terms} in the asymptotic 
      expansions of the $\iai$-interaction (\ref{interaction}) 
      and the fermionic overlap (\ref{exactom}) for large 
      conformal separation,
      \begin{equation}
      \Omega(\xi )=-\frac{6}{\xi^2}+{\mathcal O}(\ln(\xi)/\xi^4)),\hspace{6ex}
      \omega(\xi )=\frac{4}{\xi^{3/2}} + {\mathcal O}(\ln(\xi )/\xi^{7/2}),
      \label{asympt}
      \end{equation}
      {\it exactly} reproduce the known perturbative results 
      (\ref{action-asy}) and (\ref{omega-asy}) for small 
      $\rho\overline{\rho}/R^2$. 
      This illustrates the power of
      the $\iai$-valley method to effectively sum up the gluonic
      {\it final-state} tree-graph corrections to the leading semi-classical
      result\footnote{Some {\it initial-state} and {\it initial-state} - 
      final-state corrections might exponentiate as well~\cite{hard}. These are
      not taken into account by the valley action.}~\cite{holypert}.
\end{itemize}

We shall thus take the valley expressions for $\Omega$, $\omega$ and 
$\tilde\Omega$, Eqs.~(\ref{interaction}), (\ref{exactom}) and 
(\ref{orient-valley}), to smoothly 
extrapolate somewhat beyond strict $I$-perturbation theory. 

Let us add, however, that the full content of the valley approximation
is not essential in this context. It is mostly the shift in the expansion
variable 
\begin{equation}
\frac{\rho\overline{\rho}}{-R^2+\ii\epsilon R_0} \Rightarrow
\frac{\rho\overline{\rho}}{-R^2+\ii\epsilon R_0 +\rho^2
+\overline{\rho}^2}\equiv \frac{1}{\xi}
\end{equation}
making the leading terms (\ref{asympt}) qualitatively
adequate down to fairly small $\xi (\gwig 3)$, in contrast to the strict
$I$-perturbative expansions (\ref{action-asy}) and  (\ref{omega-asy}).
 
\begin{figure}[ht]
\vspace{-0.4cm}
\begin{center}
\epsfig{file=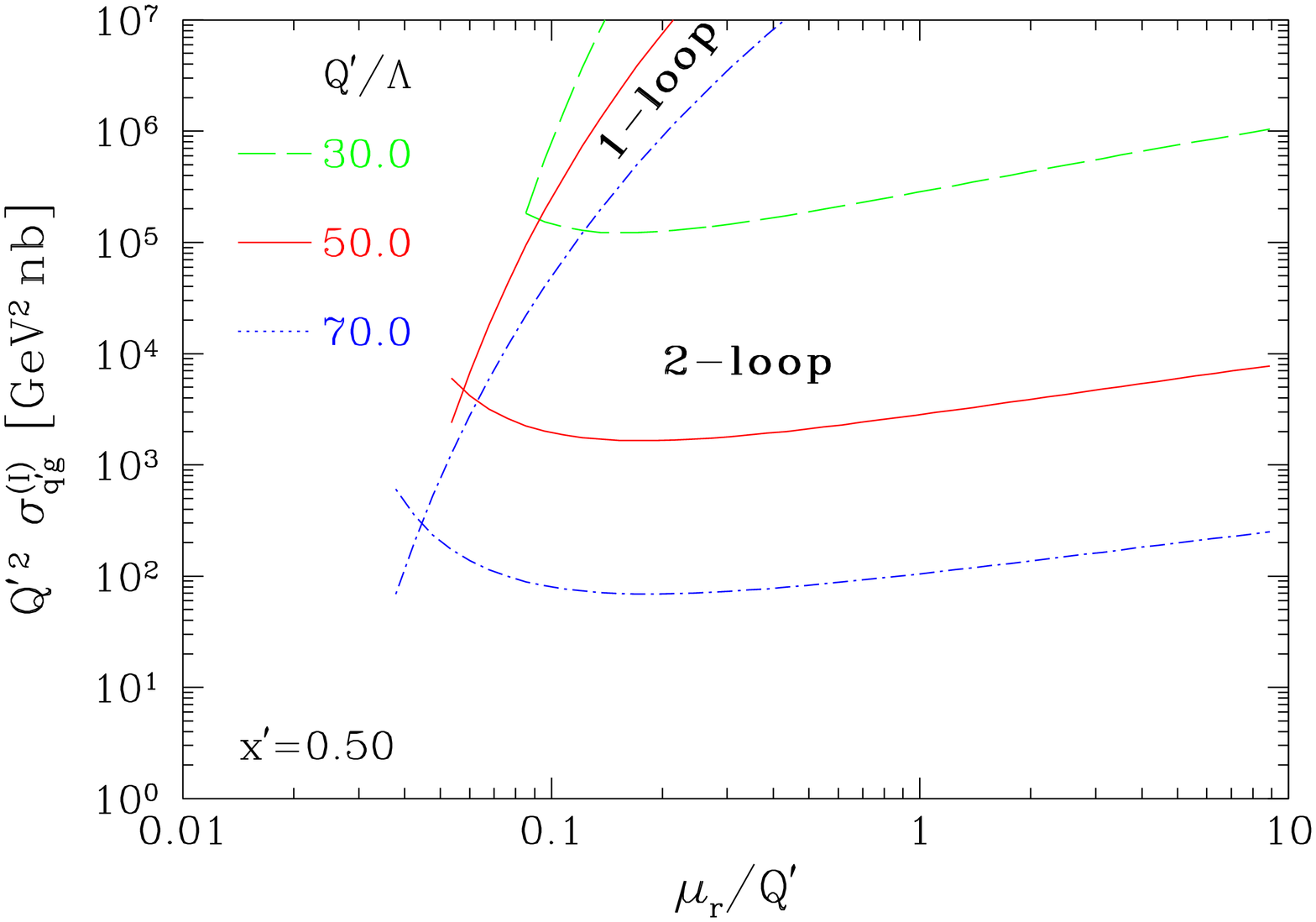,angle=270,width=8.2cm}
\epsfig{file=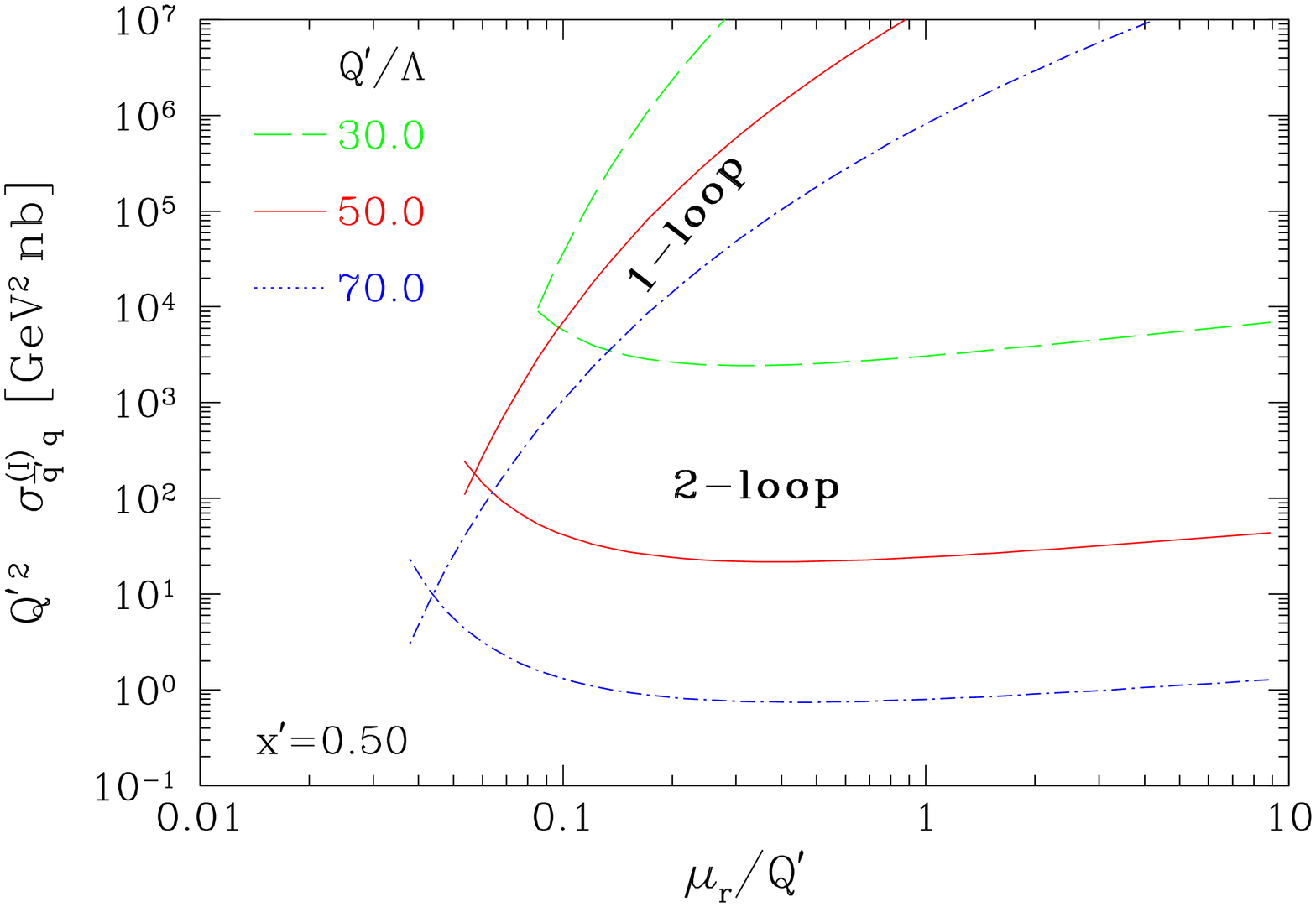,angle=270,width=8.2cm}\hfill
\caption[dum]{\label{q2sigkap}
Renormalization-scale dependences of the $I$-subprocess cross-sections
for a target gluon, Eq.~(\ref{qgcross-vxi}), (left) and a 
target quark, Eq.~(\ref{barqqcross-vxi}), both for $n_f=3$.
 }
\end{center}
\end{figure}

The collective coordinate integration in the cross-section~(\ref{qgforwampl})
is perfectly suited for a {\it saddle-point evaluation}. To this end, we 
collect the most-relevant factors in Eq.~(\ref{qgforwampl}) in the following 
effective exponent, 
\begin{eqnarray}
\label{gamma}
-\Gamma  \equiv
\ii\,(p+q^\prime)\cdot R  
-\,Q^\prime \left(\rho+\overline{\rho}\right)
-\left(\frac{4\pi}{\alpha_s\left(\mu _r \right)}-\Delta_1\,\beta_0
\ln\left(\rho\overline{\rho}\mu_r^2\right)\right) S^{(\iai )}(\xi )-
\Delta_2 \ln\left(\rho\overline{\rho}\mu_r^2\right) 
\, .
\end{eqnarray}
In arriving at Eq.~(\ref{gamma}) we have used the asymptotic form 
$K_1 (Q^\prime \rho )\propto \exp\left[- Q^\prime \rho\right]$ for the 
Bessel-K functions, anticipating that, for small $\alpha_s (\mu _r )$, the 
dominant contribution to Eq.~(\ref{qgforwampl})  
will come from the region $Q^\prime\, \rho(\overline{\rho})\gg 1$.
Note that the parameters $\Delta_1$ and $\Delta_2$ allow us to trace  
the impact of the two-loop improvement of the $I$-density, with 
the one-loop expression~\cite{th} corresponding to  
$\Delta_1=1$ and $\Delta_2=0$. This is to be contrasted with previous related 
studies which either ignored the crucial renormalization-scale
dependences altogether~\cite{valley-most-attr-orient,bbgg} or  were
still too crude for a study of the associated uncertainties~\cite{rs2} 
(c.f. also Fig.~\ref{q2sigkap} (left) below). 

The corresponding saddle-point in $R_\mu$, $\rho$ and $\overline{\rho}$ is 
most easily found in the $p^\prime g$ centre-of-mass (c.m.) system. One finds 
$R_\mu^\ast = (-\ii\rho^\ast \sqrt{\xi^\ast -2},\vec{0})$ and 
$\rho^\ast=\overline{\rho}^\ast$, where $\xi^\ast$ and $\rho^\ast$
are the solutions of the following saddle-point equations, 
\begin{eqnarray}
\label{eq1}
\frac{1}{2}\frac{\sqrt{\frac{1-\xpr}{\xpr}}}
{\sqrt{\xi^\ast -2}}\, Q^\prime\rho^\ast 
-
\left(\frac{4\pi}{\alpha_s\left(\mu _r \right)}-2\,\Delta_1\,\beta_0
\ln\left(\rho^\ast\mu_r\right)\right) 
\frac{dS^{(\iai )}(\xi_\ast )}{d\xi^\ast} 
&=&0 \,,
\\[2.4ex]
\label{eq2}
\left(
\frac{1}{2}\,\sqrt{\frac{1-\xpr}{\xpr}} \sqrt{\xi^\ast -2} 
-1\right) Q^\prime\rho^\ast+
\Delta_1\beta_0 S^{(\iai )}(\xi^\ast )
-\Delta_2
&=& 0\, .
\end{eqnarray}
Upon evaluating the integrand in Eq.~(\ref{qgforwampl}) at the saddle-point 
and taking into account the integration over the (Gaussian) fluctuations
about the saddle-point\footnote{We have checked that our result for the 
Gaussian integrations coincides, for the one-loop case ($\triangle_1=1,
\triangle_2=0$), with the corresponding result quoted in Ref.~\cite{bbgg}.}, we
may finally express the cross-section  
entirely in terms of $v^\ast\equiv Q^\prime\rho^\ast$ and $\xi^\ast$, 
\begin{eqnarray}
\label{qgcross-vxi}
Q^{\prime\,2}\,\sigma_{p^\prime g}^{(I )} &=&
d^2\frac{\sqrt{12}}{2^{16}}\pi^{15/2}
((\xi^\ast +2) v^{\ast\,2}+4 \ts (\ts-2 v^\ast))
\left( \frac{(\xi^\ast -2)}{\xi^\ast} \frac{\Delta_1\beta_0}{\dts}\right)^{7/2}
\omega (\xi^\ast )^{2n_f-1}
\\[1.6ex] &&\times
\frac{(\xi^\ast -2)^3 v^{\ast\,5}
}
{(v^\ast -\ts)^{9/2}
 \sqrt{(\xi^\ast +2)v^\ast -4\ts}
 \sqrt{\frac{1}{2}(\ts -v^\ast -2 \dts )^2 + \ts (\ts -v^\ast ) \ddts}} 
\nonumber
\\[1.6ex] &&\times
\left( \frac{4\pi}{\alpha_s\left(\mu _r \right)}\right)^{19/2}
\exp \left[ 
-\frac{4\pi}{\alpha_s\left(\mu _r \right)} S^{(\iai )}\left(\xi^\ast 
\right) - 2
\left(1-\ln\left(\frac{v^\ast\mu_r}{Q^\prime}\right)\right)
\,\ts\right] \, ,
\nonumber
\\[2.4ex] 
Q^{\prime\,2}\,\sigma_{\overline{q}^\prime q}^{(I)} &=&
\frac{32}{3}\,\frac{\alpha_s(\mu_r)}{4\pi}\,
\frac{1}{v^\ast}\,\frac{\sqrt{\xpr (1-\xpr)}}{\omega (\xi^\ast )} \,
Q^{\prime\,2}\,\sigma_{p^\prime g}^{(I )}\, ,\hspace{6ex}
\sigma_{q^\prime q}^{(I)}=(1-\delta_{q^\prime q})
\sigma_{\overline{q}^\prime q}^{(I)}\, , 
\label{barqqcross-vxi}
\end{eqnarray}
where we have introduced the shorthands
\begin{equation}
\ts (\xi^\ast )\equiv \Delta_1 \beta_0 S^{(\iai )}(\xi^\ast )-\Delta_2,
\hspace{6ex}
D(f(\xi^\ast ))\equiv 
\frac{d}{d\ln (\xi^\ast -2)}f(\xi^\ast )  .
\end{equation}
For completeness, we have listed also the corresponding expression
for the $\overline{q}^\prime q$ cross-section~(\ref{barqqcross-vxi}). 

What remains is to solve the saddle-point equations, (\ref{eq1}) and
(\ref{eq2}). An {\it analytical solution}~\cite{mrs3} in the 
asymptotic regime $\alpha_s (Q^\prime )\to 0$, $\xpr$ and $\mu_r/Q^\prime$ 
fixed, confirms our qualitative expectations (\ref{qual-exp}). 
In particular one finds, asymptotically, $R^\ast/\rho^\ast = \sqrt{\xi^\ast -2}
\to 2/\sqrt{1/\xpr -1}$. 
However, for experimentally accessible values of the 
virtuality $Q^\prime$,  the corrections to the asymptotic result are
quite large and the corresponding analytical expressions complicated. 
Hence, we only present here the results 
corresponding to a {\it numerical} solution of the saddle-point equations.

In Fig.~\ref{q2sigkap}, we display the residual renormalization-scale 
dependencies of the $I$-subprocess cross-sections~(\ref{qgcross-vxi}) and 
(\ref{barqqcross-vxi}) over a {\it large} range of $\mu_r/Q^\prime$. 
Apparently, we have achieved great progress 
in stability and hence predictivity by using the two-loop 
renormalization-group invariant form of the 
$I$-density $D(\rho,\mu_r)$ from Eqs.~(\ref{density}) and (\ref{Deltas}): 
The residual dependence on the renormalization scale $\mu_r$ turns out to
be strongly reduced as compared to the one-loop case ($\Delta_1=1,\Delta_2=0$).

Intuitivelely one may expect~\cite{mrs1,bbgg,bb} $\mu_r \sim 1/\langle \rho
\rangle \sim Q^\prime/\beta_0 ={\cal O}(0.1)\, Q^\prime$. Indeed, 
this guess turns out to match quite well
our actual choice of the  
 ``best'' scale, $\mu_r = 0.15\ Q^\prime$, for which 
$\partial \sigma^{(I)}_{q^\prime g}/\partial \mu_r \simeq 0$
(c.f. Fig.~\ref{q2sigkap} (left)). 
We also note that the 
cross-sections for a target gluon (Fig.~\ref{q2sigkap} (left)) are about two 
orders of magnitudes larger than the cross-sections for a target quark 
(Fig.~\ref{q2sigkap} (right)). Henceforth, the latter are neglected.  

\begin{figure}[ht]
\vspace{-0.4cm}
\begin{center}
  \epsfig{file=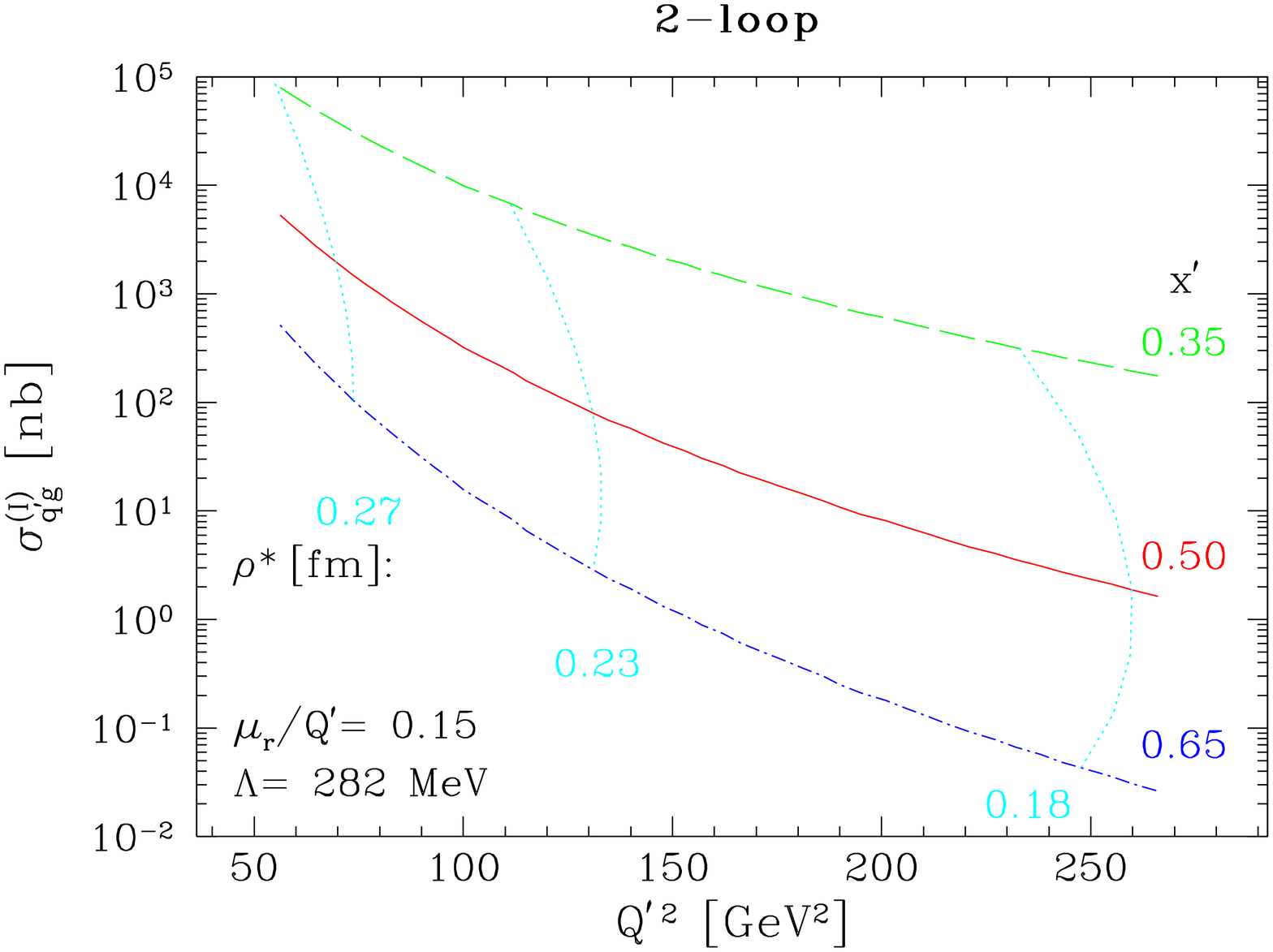,angle=270,width=8.2cm}
  \epsfig{file=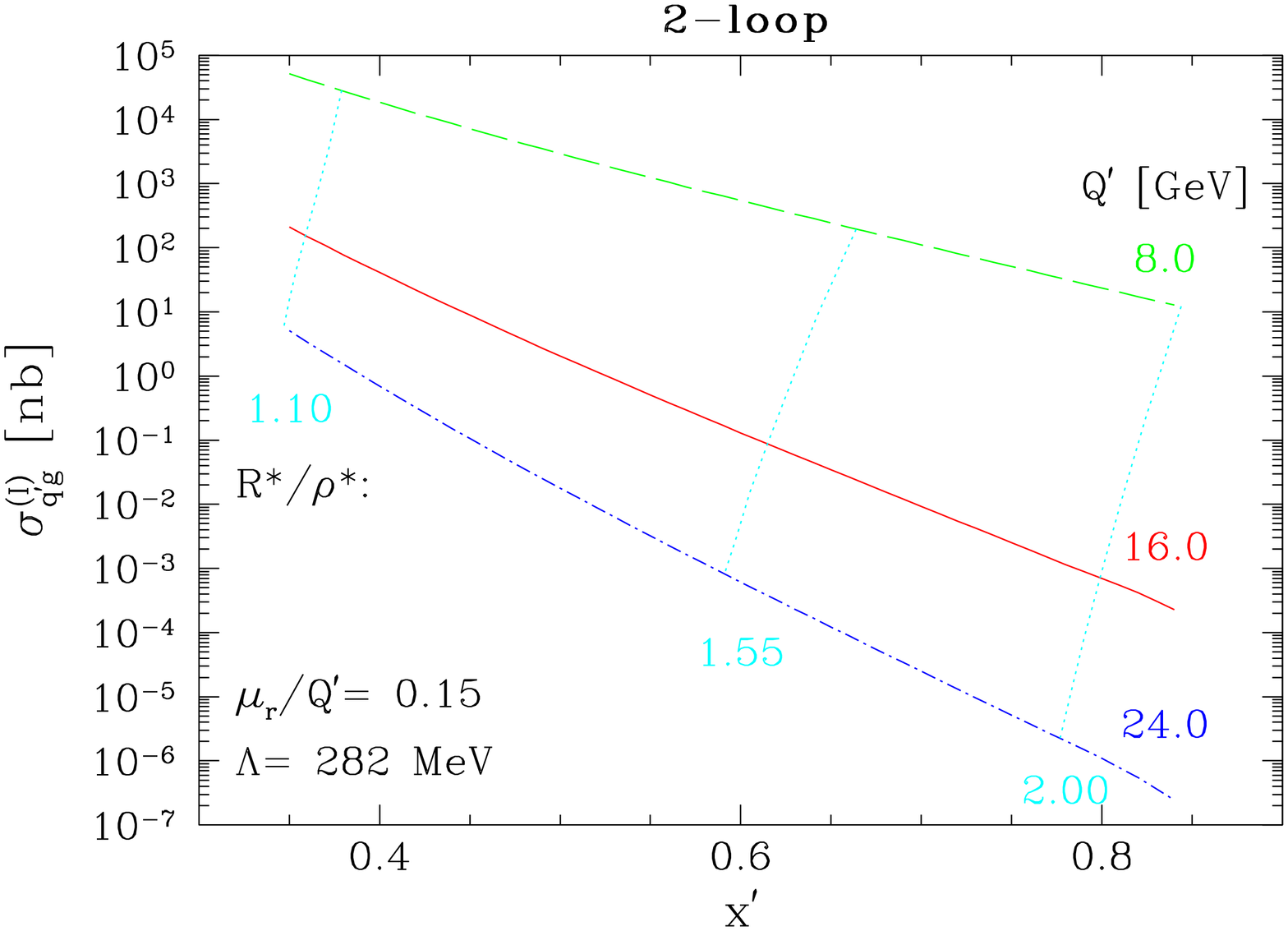,angle=270,width=8.2cm}
\caption[dum]{\label{isorho}
The instanton-subprocess cross-section~(\ref{qgcross-vxi}), for $n_f=3$, both 
as functions of $Q^{\prime 2}$ (left) and $\xpr$ (right). The dotted curves 
are lines of constant $I$-size $\rho^\ast$ (left) and of constant 
$\iai$-separation $R^\ast$ in units of the $I$-size $\rho^\ast$ 
(right). 
 }
\end{center}
\end{figure}
Our quantitative results on $\sigma^{(I)}_{q^\prime g}$ are shown in detail 
in Fig.~\ref{isorho}, both as functions of $Q^{\prime 2}$ (left) and of $\xpr$
(right). The dotted curves indicating lines of constant $\rho^\ast$ (left)
and of constant $R^\ast/\rho^\ast$ (right)    
nicely illustrate the qualitative relations~(\ref{qual-exp}):  
For growing $Q^{\prime 2}$ and fixed $\xpr$, smaller and smaller instantons,
$\rho^\ast\sim 1/Q^\prime$,  
are probed and the cross-sections decrease rapidly, mainly because of 
the large powers of $\rho$ in the $I$-density~(\ref{density}).  
For decreasing $\xpr$ and fixed $Q^{\prime 2}$, on the other hand, the 
$\iai$-separation $R^\ast$ in units of the $I$-size
$\rho^\ast$ decreases and the cross-section increases dramatically. In the 
language of the $\iai$-valley method the latter originates mainly from the
attractive interaction between instantons and anti-instantons.


{\bf 3.}
We have seen that the collective coordinate integrals in (\ref{qgforwampl}) 
are dominated by a single, calculable saddle-point
($\rho^\ast,R^\ast/\rho^\ast$), in one-to-one
relation to the conjugate momentum variables ($Q^\prime ,\xpr$). 
This effective one-to-one
mapping of the conjugate $I$-variables allows for the following 
important strategy:
We may determine {\it quantitatively} the range of validity
of $I$-perturbation theory and the dilute $I$-gas approximation in the 
instanton collective coordinates ($\rho\leq \rho_{\rm max},
R/\rho\geq (R/\rho)_{\rm min}$) from recent (non-perturbative) lattice
simulations of QCD and translate the resulting constraints via the
mentioned one-to-one relations into a ``fiducial'' kinematical region 
($Q^\prime\geq Q^\prime_{\rm min},\xpr\geq x^\prime_{\rm min}$).

In lattice simulations 4d-Euclidean space-time is made discrete;
specifically, the ``data'' from the UKQCD
collaboration~\cite{ukqcd}, which we shall use here, involve a lattice
spacing $a = 0.055 - 0.1$ fm and a volume  $V=l_{\rm space}^{\,3}\cdot l_{\rm time}=[16^3\cdot 48 - 32^3\cdot 64]\,a^4$.  
In principle, such a lattice allows to study the properties of an 
ensemble of (anti-)instantons  with sizes $a < \rho < V^{1/4}$. However,
in order to make instanton effects visible, a certain ``cooling'' 
procedure has to be applied first. It is designed to
filter out (dominating) fluctuations of {\it short} wavelength
${\cal O}(a)$, while affecting the topological fluctuations of much longer
wavelength $\rho \gg a$ comparatively little. For a discussion of 
lattice-specific caveats, like possible lattice artefacts and the
dependence of results on ``cooling'' etc., see Refs.~\cite{lattice,ukqcd}.   

\begin{figure}[ht]
\vspace{-0.8cm}
\begin{center}
\epsfig{file=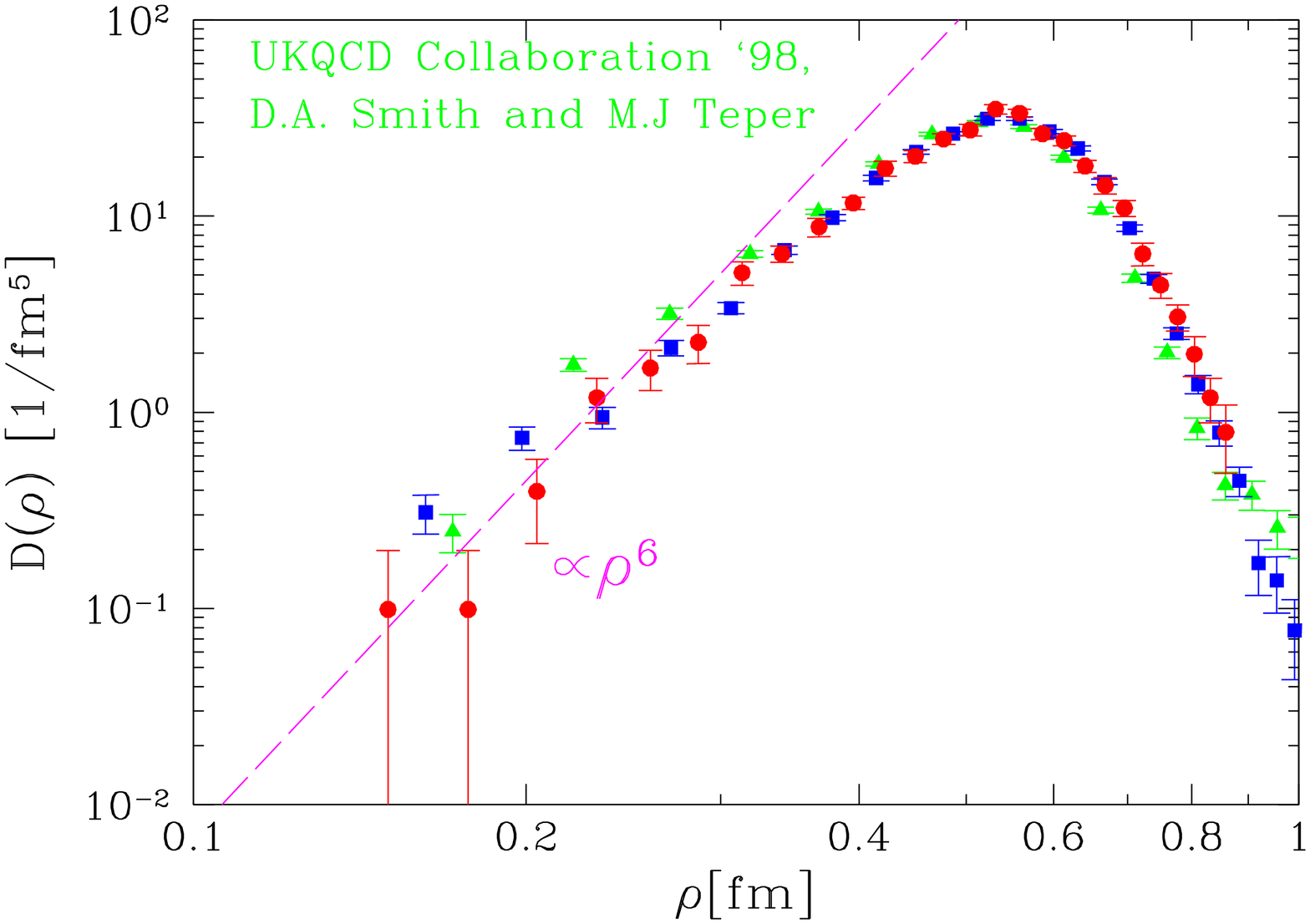,angle=-90,%
width=8.2cm}
\epsfig{file=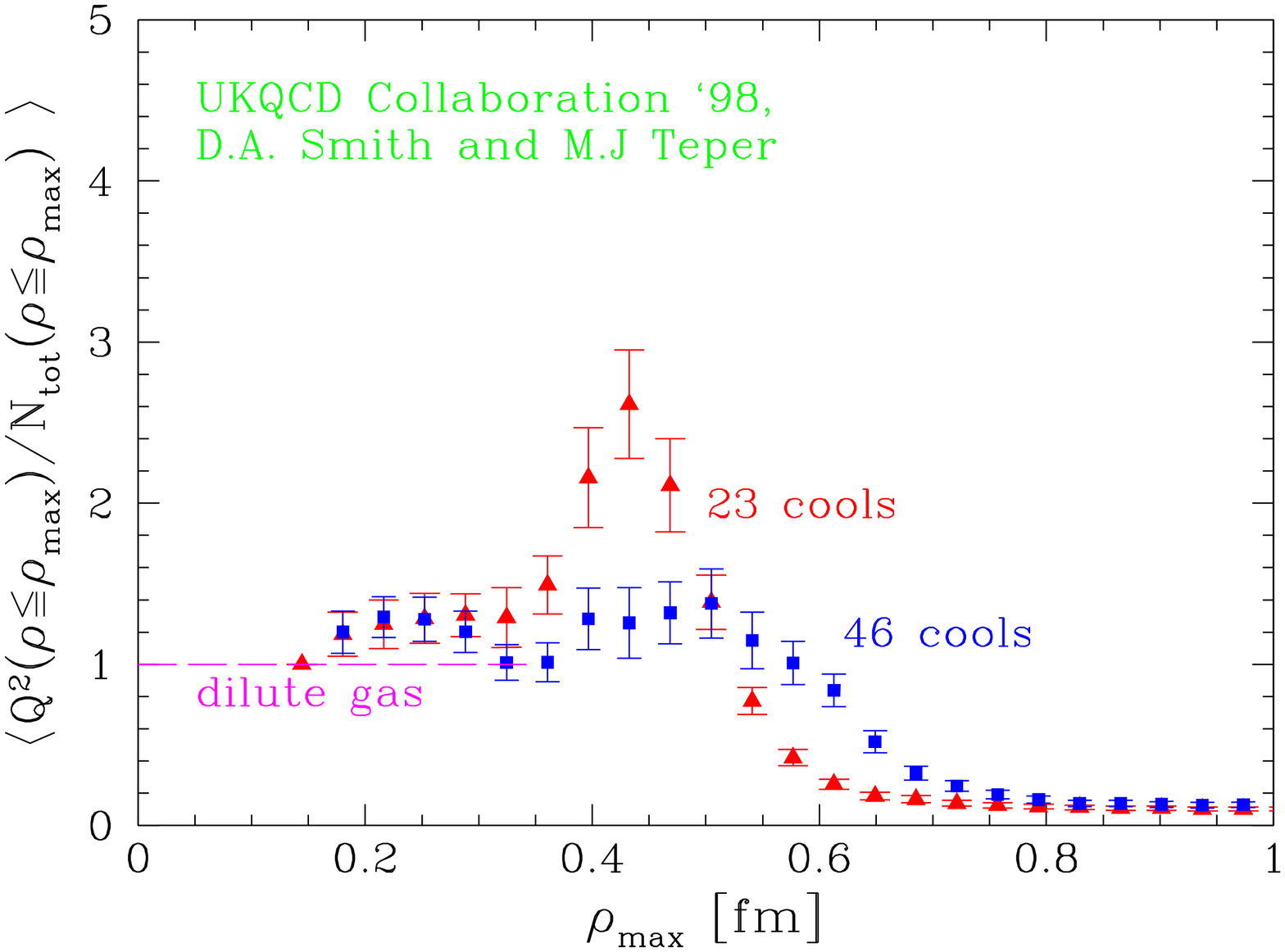,angle=-90,%
width=8.2cm}
\caption[dum]{\label{lattice-support}Support for the validity of 
$I$-perturbation theory for the
$I$-density $D(\rho)$  (left) and the dilute $I$-gas approximation
(right) for $\rho<\rho_{\rm max}\simeq 0.3$ fm from recent lattice
data~\cite{ukqcd}.
 }
\end{center}
\end{figure}
The first important quantity of interest, entering $I$-induced cross-sections 
(c.f.~Eq.~(\ref{qgforwampl})), is 
the {\it $I$-density} $D(\rho)$, Eq.~(\ref{density}). This power law,
$D(\rho)_{\mid n_f=0}\propto \rho^6$, of
$I$-perturbation theory is confronted in Fig.~\ref{lattice-support}\,(left) 
with recent lattice ``data'',
which strongly suggests semi-classical $I$-perturbation theory to be valid for
$\rho \lwig \rho_{\rm max}\simeq 0.3$ fm. 
Next, consider the square of the total topological charge,
$Q^2=(n-\bar{n})^2$, along with the
total number of charges, $N_{\rm tot}=n+\bar{n}$.
For a {\it dilute gas}, the number fluctuations are {\it poissonian} and 
correlations among the $n$ and $\bar{n}$ distributions
absent, implying 
$\langle Q^2/N_{\rm tot} \rangle =1$.
From Fig.~\ref{lattice-support}\,(right), it is apparent that this relation,
characterizing the validity of the dilute $I$-gas approximation, is well 
satisfied for sufficiently {\it small} instantons. Again, we find 
$\rho_{\rm max}\simeq 0.3$ fm, quite independent of the number of
cooling sweeps. For increasing  $\rho_{\rm max}\gwig 0.3$ fm, the ratio 
$\langle Q^2/N_{\rm tot}\rangle$ rapidly and strongly deviates
from one.

Crucial information about a second quantity of interest, the 
{\it $\iai$-interaction}, may be obtained as well from the 
lattice~\cite{lattice,ukqcd}. Quite generally, it is found that the 
semi-classical attraction for large $R^2/(\rho\overline{\rho})$ 
turns into a non-perturbative repulsion for smaller separations in units of
the sizes, such that in vacuum\footnote{Published ratios range from 
$\langle R\rangle/\langle \rho \rangle\simeq 0.83$~\cite{ukqcd}, 
$\langle R/(\rho +\overline{\rho})\rangle \simeq 0.59$~\cite{fps} to
$\langle R/(\rho +\overline{\rho})\rangle \simeq 1$~\cite{dhk}.}
 $\langle R^2/(\rho\overline{\rho})\rangle =
{\mathcal O}(1)$. Thus it seems a reasonable extrapolation to use
the attractive, semi-classical valley result for the $\iai$-interaction
$\Omega$, Eq.~(\ref{interaction}), down to a minimum conformal separation 
$\xi_{\rm min}\simeq 3$, corresponding to 
$(R^\ast/\rho^\ast)_{\rm min}\simeq 1$.

Finally, by means of the discussed 
saddle-point translation, these lattice constraints may be turned 
into a ``fiducial'' kinematical region for our 
cross-section predictions in DIS (c.f. Fig.~\ref{isorho}),
\begin{equation}
 \left.\begin{array}{lcccl}\rho^\ast&\leq& \rho^\ast_{\rm max}&\simeq& 
         0.3 {\rm\ fm};\\[1ex]
 \frac{R^\ast}{\rho^\ast}&\geq&\left(\frac{R^\ast}{\rho^\ast}\right)_{\rm min}
 &\simeq& 1\\
 \end{array}\right\}\Rightarrow
 \left\{\begin{array}{lclcl}Q^\prime&\geq &Q^\prime_{\rm min}&\simeq&
 8 {\rm\ GeV};\\[1ex]
 x^\prime&\geq &x^\prime_{\rm min}&\simeq &0.35.\\
 \end{array} \right .
\label{fiducial}
\end{equation}

Unlike DIS, where only {\it small} instantons are probed, 
in the $I$-liquid model of Ref.~\cite{shuryak} more emphasis is placed
on the physics associated with {\it larger} instantons. For $I$-ensembles
including also larger $I$-sizes $\gwig 0.3$ fm, the various recent lattice 
results~\cite{lattice,ukqcd,fps,dhk} do not, however, unanimously support 
the liquid picture.   


{\bf 4.}
Experimentally, in deep inelastic $eP$ scattering at HERA, the 
cuts~(\ref{fiducial}) must be implemented via a ($Q^\prime ,\xpr$) 
reconstruction from the final-state momenta and topology~\cite{crs}, 
while theoretically, they are incorporated into our $I$-event 
generator~\cite{grs} ``QCDINS 1.6.0'' and the resulting prediction of 
the $I$-induced cross-section in DIS at HERA.
The latter is connected to the $I$-subprocess cross-sections 
$\sigma^{(I)}_{p^\prime\,p}$ by the differential
$p^\prime p$ luminosity~\cite{mrs3} (c.f. Eq.~(\ref{ePcross})), 
\begin{eqnarray}
\label{diff-lum}
\frac{d{\mathcal L}^{(I)}_{p^\prime p}}{d\xpr\,dQ^{\prime 2}}
=\frac{2\pi\alpha^2}{S}\frac{e_{p^\prime}^2}{\xpr^2}
\int\limits_{x_{\rm Bj\,min}}^\xpr \frac{dx}{x} 
\int\limits_{x_{\rm Bj\,min}}^x \frac{dx_{\rm Bj}}{x_{\rm Bj}}
\int\limits_{y_{\rm Bj\,min}}^{y_{\rm Bj\,max}} \frac{dy_{\rm Bj}}{y_{\rm Bj}}
\,P_{\gamma^\ast}(y_{\rm Bj})\,
P_{p^\prime}^{(I)}(\frac{x}{\xpr},\ldots )\,
f_p (\frac{x_{\rm Bj}}{x},\ldots ) .
\end{eqnarray}  
Here $S\,(\simeq 9\cdot 10^4\ {\rm GeV}^2$ for HERA) denotes the c.m. energy 
squared of the $eP$ collision, $e_{p^\prime}^2$ is the electric charge squared
of the current (anti-)quark in units of the electric charge squared,
$e^2=4\pi\alpha$, and $P_{\gamma^\ast}$ denotes the familiar
Weizs\"acker-Williams-type photon flux,
\begin{eqnarray}
P_{\gamma^\ast}(y_{\rm Bj}) = 
(1+(1-y_{\rm Bj})^2)/y_{\rm Bj}, 
\end{eqnarray}
with $y_{\rm Bj}=Q^2/(S x_{\rm Bj})$. Furthermore, 
$f_p (x_{\rm Bj}/x,\ldots )$ denotes the density of the target parton 
$p$ in the proton, with the dots standing for the factorization scale, and,
finally, the factor $P_{p^\prime}^{{ (I)}}$ accounts for the flux of 
virtual (anti-)quarks $p^\prime$ in the $I$-background entering the 
$I$-induced $p^\prime p$-subprocess from the photon side~\cite{rs2,mrs3} (c.f. 
Fig.~\ref{ev-displ}), 
\begin{eqnarray}
P_{q^\prime}^{{ (I)}}\,\left( 
\frac{x}{x^\prime},x ,\frac{Q^{\prime}}{Q}\right)\ \equiv 
P_{\overline{q}^\prime}^{{ (I)}}\,\left( 
\frac{x}{x^\prime},x ,\frac{Q^{\prime}}{Q}\right)\
\simeq \frac{3}{16\,\pi^3}\,\frac{x}{x^\prime}
\left(1+\frac{1}{x}-\frac{1}{x^{\prime}}-\frac{Q^{\prime 2}}{Q^{2}}\right)
\, .
\end{eqnarray}

The $I$-induced cross-section in DIS at HERA, $\sigma^{(I)}_{\rm HERA}$, 
subject to kinematical cuts 
($x_{\rm Bj}\geq x_{\rm Bj\,min}$; $y_{\rm Bj\,max}\geq y_{\rm Bj}\geq 
y_{\rm Bj\,min}$; $\xpr\geq x^\prime_{\rm min}$; $Q^\prime\geq 
Q^\prime_{\rm min}$), 
is then obtained by integrating Eq.~(\ref{ePcross}) over the appropriate 
range of $\xpr$ and $Q^{\prime 2}$.  

Let us point out that the factorization of the $eP$ cross-section
for fixed  $\xpr$ and $Q^{\prime 2}$ into a sum of differential 
luminosities and $I$-subprocess cross-sections, is essential for 
the possibility to place {\it different cuts} on the Bjorken 
variables of the $eP$ and the $p^\prime p$ system, respectively. Of particular
interest is $x_{\rm Bj\,min}\ll x^\prime_{\rm min}$. Such cuts permit to 
explore  essentially the full accessible ($x_{\rm Bj},x,Q^2$) 
range in DIS at HERA, down to ($10^{-3},10^{-3},10$ GeV$^2$), say. 
By placing in this region the additional cuts~(\ref{fiducial}) 
on ($\xpr,Q^{\prime 2}$), $I$-searches benefit from the high
statistics at small $x_{\rm Bj}$, while the theoretical control is
retained over the $I$-dynamics. 

In  Ref.~\cite{bb}, on the other hand,  (only) the 
{\it infrared safe} pieces of the $I$-induced contributions to the 
parton-structure functions, ${\mathcal F}^{(I)}_{2\,p}(x,Q^2)$, were 
estimated in one step by means of configuration space techniques.
In our momentum space language, the authors have implicitly {\it integrated} 
over $Q^{\prime 2}$ and $\xpr$, with $x_{\rm Bj}\le x\le \xpr\le 1$.
Hence, the results of Ref.~\cite{bb} can only be applied~\cite{rs} to
relatively large  $x_{\rm Bj\,min}=x^\prime_{\rm min}\sim 0.35$. 

Nevertheless, as an important check of our calculations, we have also 
calculated the {\it infrared-safe}, $I$-induced contributions to the 
parton-structure functions, by integrating our asymptotic results 
over $\xpr$ and $Q^{\prime 2}$ and retaining only the contributions
from the upper  $Q^{\prime 2}$-integration limit $\propto Q^2$. 
Within the common range of validity~\cite{mrs3} of various employed
approximations, we find perfect agreement with the 
gluon-structure function quoted in Ref.~\cite{bb}.

\begin{figure}[ht]
\vspace{-0.1cm}
\begin{center}
  \epsfig{file=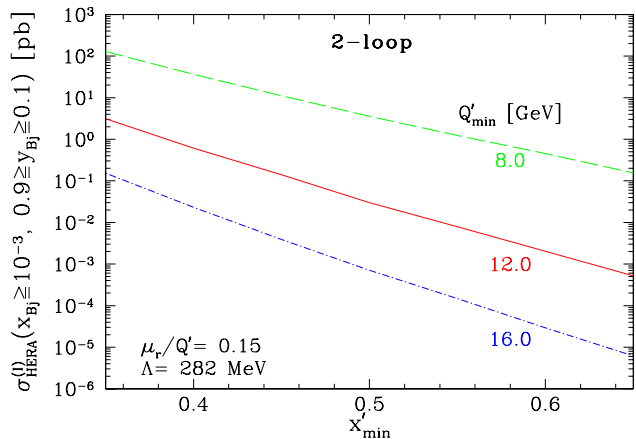,angle=270,width=9.5cm}
\caption[dum]{\label{sigHERA}Instanton-induced cross-section at HERA 
($n_f=3$).}
\end{center}
\end{figure}
Fig.~\ref{sigHERA} displays the finalized $I$-induced cross-section at HERA,
as function of the cuts $x^\prime_{\rm min}$ and $Q^\prime_{\rm min}$, as
obtained with the new release ``QCDINS 1.6.0'' of our $I$-event
generator. Only the target gluon contribution has been taken into account.
For the minimal cuts 
(\ref{fiducial}) extracted from lattice simulations, we
specifically obtain 
\begin{equation}
\label{minimal-cuts}
\sigma^{(I)}_{\rm HERA}(\xpr\ge0.35,Q^\prime\ge 8\, {\rm GeV}) 
\simeq 126\, {\rm pb};\hspace{4ex}
 {\rm for}\ x_{\rm Bj}\ge 10^{-3};\ 0.9\ge y_{\rm Bj}\ge 0.1 .
\end{equation}
Hence, with the total luminosity accumulated by experiments at HERA, 
${\mathcal L}={\mathcal O}(80)$ pb$^{-1}$, there should be already 
${\mathcal O}(10^4)$ $I$-induced events from this kinematical region
on tape. 
Note also that the cross-section quoted in Eq.~(\ref{minimal-cuts})
corresponds to a fraction of $I$-induced to normal DIS (nDIS) events of
\begin{equation}
f^{(I)} = \frac{\sigma^{(I)}_{\rm HERA}}{\sigma^{({\rm nDIS})}_{\rm HERA}}
={\mathcal O}(1)\, \%; 
\hspace{6ex} {\rm for}\ x_{\rm Bj}\ge 10^{-3};\ 0.9\ge y_{\rm Bj}\ge 0.1 .
\end{equation}
This is remarkably close to the published upper limits on the fraction of
$I$-induced events~\cite{limits}, which are also on the one percent
level.

There are still a number of significant uncertainties in our cross-section 
estimate. 
For {\it fixed $Q^\prime$- and $\xpr$-cuts}, one of the dominant uncertainties
arises from the experimental uncertainty in the QCD scale $\Lambda$. We used 
in the two-loop expression for $\alpha_s$ with $n_f=3$ massless flavours the 
value $\Lambda_{\overline{\rm MS}}^{(3)}=282$ MeV, corresponding to the 
central value of the DIS average for $n_f=4$, 
$\Lambda_{\overline{\rm MS}}^{(4)}=234$ MeV~\cite{pdg}. If we change 
$\Lambda_{\overline{\rm MS}}^{(3)}$ within the allowed range, 
$\approx\pm 65$ MeV, the cross-section (\ref{minimal-cuts}) varies
between 26 pb and 426 pb. Minor uncertainties are associated with  
the residual renormalization-scale dependence (c.f. Fig.~\ref{q2sigkap}) and 
the choice of the factorization scale. Upon varying the latter by an order of 
magnitude, the changes are in the ${\mathcal O}(20)$ \% range.

By far the most dominant uncertainty arises, however, from the unknown 
boundaries of the fiducial region in ($\xpr ,Q^\prime$) 
(c.f. Fig.~\ref{sigHERA}). Here, the constraints from lattice simulations are 
extremely valuable for making concrete predictions.



\vspace{0.2cm}
\section*{Acknowledgements}
We would like to thank S. Moch for valuable contributions in the early stage
of this work. 
We would like to acknowledge helpful discussions with I. Balitsky,  
V. Braun and N. Kochelev.


\vspace{0.2cm}

\begin{thebibliography}{99}
\bibitem{bpst}
A. Belavin, A. Polyakov, A. Schwarz and Yu. Tyupkin, 
\Journal{\PLB}{59}{85}{1975}.
\bibitem{th} 
G. `t Hooft, \Journal{\PRL}{37}{8}{1976};
\Journal{\PRD}{14}{3432}{1976}; \Journal{\PRD}{18}{2199}{1978}
(Erratum).
\bibitem{bb} 
I. Balitsky and V. Braun, \Journal{\PLB}{314}{237}{1993}. 
\bibitem{mrs1}
S. Moch, A. Ringwald and F. Schrempp, 
\Journal{\NPB} {507}{134}{1997}.
\bibitem{rs}
A. Ringwald and F. Schrempp, hep-ph/9411217, in:
{\it Quarks `94}, Proc. 8th Int. Seminar, Vladimir, Russia, 
May 11-18, 1994, eds. D. Gigoriev et al., pp. 170-193.
\bibitem{grs}
M. Gibbs, A. Ringwald and F. Schrempp, hep-ph/9506392,
in: {\it Proc. Workshop on Deep Inelastic Scattering and QCD},
Paris, France, April 24-28, 1995, eds. J.-F. Laporte and Y. Sirois,
pp. 341-344.
\bibitem{rs2}
A. Ringwald and F. Schrempp, hep-ph/9610213,
in: 
{\it Quarks `96}, Proc. 9th Int. Seminar, Yaroslavl, Russia, May 5-11, 1996,
eds. V. Matveev, A. Penin, V. Rubakov, A. Tavkhelidze, Vol. I, 
pp. 29-54.
\bibitem{crs}
T. Carli, A. Ringwald and F. Schrempp, in preparation.
\bibitem{mrs3}
S. Moch, A. Ringwald and F. Schrempp, in preparation. 
\bibitem{morretal}
T. Morris, D. Ross and C. Sachrajda, 
\Journal{\NPB}{255}{115}{1985}.
\bibitem{lattice}
For a recent review, see:
P. van Baal, hep-lat/9709066, Review at Lattice `97, 
Edinburgh.
\bibitem{ber}
C. Bernard, \Journal{\PRD}{19}{3013}{1979}.
\bibitem{dMS}
A. Hasenfratz and P. Hasenfratz, \Journal{\NPB}{193}{210}{1981};
\hfill\break
M. L\"uscher, \Journal{\NPB}{205}{483}{1982};
\hfill\break
G. `t Hooft, \Journal{\PR}{142}{357}{1986}.
\bibitem{b}
I. Balitsky, hep-ph/9405335, in: {\it Continuous advances in QCD}, Proc.,  
Minneapolis, USA, Feb. 18-20, 1994, ed. A. Smilga, pp.167-194. 
\bibitem{valley-most-attr-orient}
V.V. Khoze and A. Ringwald, 
\Journal{\PLB}{259}{106}{1991}.
\bibitem{bbgg}
I. Balitsky and V. Braun,
\Journal{\PRD}{47}{1879}{1993}.
\bibitem{holypert}
P. Arnold and M. Mattis,
\Journal{\PRD}{44}{3650}{1991};
\hfill\break
A. Mueller,
\Journal{\NPB}{364}{109}{1991};
\hfill\break
D. Diakonov and V. Petrov, in:
{\it Proc. of the 26th LNPI Winter School},
(Leningrad, 1991), pp. 8-64.
\bibitem{optvalley}
M. Porrati,
\Journal{\NPB}{347}{371}{1990};
\hfill\break
V.V. Khoze and A. Ringwald,
\Journal{\NPB}{355}{351}{1991};
\hfill\break
V. Zakharov,
\Journal{\NPB}{371}{637}{1992};
\hfill\break
I. Balitsky and V. Braun,
\Journal{\NPB}{380}{51}{1992}.
\bibitem{yung}
A. Yung,
\Journal{\NPB}{297}{47}{1988}.
\bibitem{valley-gen-orient}
J. Verbaarschot, 
\Journal{\NPB}{362}{33}{1991}.
\bibitem{shuverb}
E. Shuryak and J. Verbaarschot, 
\Journal{\PRL}{68}{2576}{1992}. 
\bibitem{cdg}
C. Callan, R. Dashen and D. Gross, 
\Journal{\PRD}{17}{2717}{1978}.
\bibitem{hard}
A. Mueller, 
\Journal{\NPB}{348}{310}{1991}; 
\Journal{\NPB}{353}{44}{1991}.
\bibitem{ukqcd}
D. Smith and M. Teper, Edinburgh preprint 98-1, hep-lat/9801008.
\bibitem{fps}
P. de Forcrand, M. P\'erez and I. Stamatescu,
\Journal{\NPB}{499}{409}{1997}. 
\bibitem{dhk}
T. DeGrand, A. Hasenfratz and T. Kov\'acs,
\Journal{\NPB}{505}{417}{1997}.
\bibitem{shuryak}
E.V. Shuryak,
\Journal{\NPB}{198}{83}{1982};
\hfill\break
T. Sch\"afer and E.V. Shuryak,
\Journal{\em Rev. Mod. Phys.}{70}{323}{1998}.
\bibitem{pdg}
Review of Particle Physics, Particle Data Group, 
\Journal{\PRD}{54}{1}{1996}.
\bibitem{limits}
S. Aid {\it et al.}, H1 Collaboration, \Journal{\NPB} {480}{3}{1996};
\hfill\break
S. Aid {\it et al.}, H1 Collaboration, \Journal{\ZPC} {72}{573}{1996};
\hfill\break
T. Carli and M. Kuhlen, \Journal{\NPB} {511}{85}{1998}.


\end{thebibliography}
\end{document}